\documentclass[aps,prl,preprint]{revtex4-1}

\usepackage{comment, color, graphicx, amsmath}

\newcommand{\vect}[1]{\mbox{\boldmath $#1$}}
\newcommand{\tens}[1]{\mbox{\boldmath $\overleftrightarrow{#1}$}}
\newcommand{\FP}{Fokker-Planck~}
\newcommand{\Cl}{C_{\mathrm{i}}}
\newcommand{\Phio}{\Phi_0}
\newcommand{\vv}{v}

\newcommand{\sigmaneo}{\sigma_{\mathrm{neo}}}
\newcommand{\CI}{C_{\mathrm{I}}}
\newcommand{\vi}{\vv_{\mathrm{i}}}
\newcommand{\fee}{f_{\mathrm{e}}}
\newcommand{\fii}{f_{\mathrm{i}}}
\newcommand{\omegat}{\omega_{\mathrm{t}}}

\newcommand{\fMe}{f_{\mathrm{Me}}}
\newcommand{\fMi}{f_{\mathrm{Mi}}}
\newcommand{\Bp}{B_{\theta}}
\newcommand{\gradv}{\nabla_{\vect{v}}}
\newcommand{\jpar}{j_{||}}

\newcommand{\Bav}{B_{\mathrm{av}}}
\newcommand{\Kloc}{K_{\mathrm{L}}}
\newcommand{\Knl}{K_{\mathrm{NL}}}

\newcommand{\fe}{f_{\mathrm{e}}}

\newcommand{\Te}{T_{\mathrm{e}}}
\newcommand{\Ti}{T_{\mathrm{i}}}
\newcommand{\nee}{n_{\mathrm{e}}}
\newcommand{\pe}{p_{\mathrm{e}}}
\newcommand{\ppi}{p_{\mathrm{i}}}
\newcommand{\Lone}{\mathcal{L}_{31}}
\newcommand{\Ltwo}{\mathcal{L}_{32}}

\newcommand{\LTi}{\mathcal{L}_{\Ti}}
\newcommand{\LnT}{\mathcal{L}_{nT}}

\newcommand{\Ce}{C_{\mathrm{e}}}
\newcommand{\Cee}{C_{\mathrm{ee}}}

\newcommand{\Cei}{C_{\mathrm{ei}}}
\newcommand{\nuei}{\nu_{\mathrm{ei}}}
\newcommand{\gradpar}{\nabla_{||}}

\newcommand{\rhop}{\rho_\theta}
\newcommand{\kV}{k_{||}}

\newcommand{\me}{m_{\mathrm{e}}}
\newcommand{\mi}{m_{\mathrm{i}}}
\renewcommand{\ni}{n_{\mathrm{i}}}
\newcommand{\nustar}{\nu_*}

\newcommand{\Epar}{E_{||}}
\newcommand{\vpar}{\vv_{||}}

\newcommand{\Vipar}{V_{\mathrm{i}||}}

\newcommand{\vme}{\vect{v}_{\mathrm{me}}}
\newcommand{\vE}{\vect{v}_E}
\newcommand{\vEo}{\vect{v}_{E0}}
\newcommand{\vEone}{\vect{v}_{E1}}

\newcommand{\sgn}{\mathrm{sgn}}

\newcommand{\be}{\begin{equation}}
\newcommand{\ee}{\end{equation}}

\newcommand{\vd}{\vect{v}_{\mathrm{d}}}
\newcommand{\vdo}{\vect{v}_{d0}}
\newcommand{\vm}{\vect{v}_{\mathrm{m}}}

\newcommand{\PS}{Pfirsch-Schl\"{u}ter~}

\begin{document}


\title{Changes to neoclassical flow and bootstrap current in a tokamak pedestal}



\author{Matt Landreman}
\email[]{landrema@mit.edu}
\author{Darin R. Ernst}
\affiliation{Plasma Science and Fusion Center, MIT, Cambridge, MA, 02139, USA}


\date{\today}

\begin{abstract}

In a tokamak pedestal, radial scale lengths can become comparable
to the ion orbit width, invalidating conventional neoclassical
calculations of flow and bootstrap current.  In this work we illustrate a
non-local approach that allows strong radial density variation while
maintaining small departures from a Maxwellian distribution.
Non-local effects 
alter the magnitude and poloidal variation of the flow and current.  The approach is
implemented in a new global $\delta f$ continuum code
using the full linearized Fokker-Planck collision operator. Arbitrary collisionality and aspect ratio are allowed
as long as the poloidal magnetic field is small compared to the total magnetic field.  Strong radial
electric fields, sufficient to electrostatically confine the ions, are
also included.
These effects may be important to
consider in any comparison between experimental pedestal flow
measurements and theory.

\end{abstract}


\maketitle 



In the H-mode edge pedestal of a tokamak,
strong density and temperature gradients drive large a neoclassical flow and bootstrap current.
This flow and current affect stability of the region to ELMs and other modes.
However, conventional neoclassical calculations are invalid in the pedestal since they
rely on an expansion\cite{HintonHazeltine,PerBook} in the smallness of the poloidal ion gyroradius $\rhop$ to the perpendicular scale length of density and temperature $r_\bot$.
In the pedestal, this ratio $\rhop/r_\bot$ is not small.
(We do not claim $r_\bot$ scales with $\rhop$, only that the lengths happen to be comparable in existing devices.)
Physically, conventional neoclassical theory is based upon the smallness of the orbit width ($\sim \rhop$ for ions) relative to
equilibrium profiles, yielding a local theory: flows and fluxes on one flux surface are determined by
values and gradients of pressure $p$ and temperature $T$ and the electric field at that flux surface only.
In the pedestal, however,
equilibrium profiles can vary strongly on scale of the ion orbit width,
requiring a global (nonlocal) calculation that does not rely on
the conventional $\rho_{\theta}/r_{\perp}$ expansion.

In this work, we generalize neoclassical calculations both analytically and numerically to the case
of a strong density pedestal (with density scale-length $r_n \sim \rhop$) as long as the ion temperature scale length
$r_T$ remains $\gg \rhop$, with a few other assumptions.
We demonstrate how the neoclassical flow is altered,
and the resulting poloidal flow variation will be important to consider for understanding experimental pedestal flow measurements.
More generally, we emphasize that compared to the general $r_T \sim \rhop$ case, this ``weak-$\Ti'$ pedestal"
is much more amenable to analysis: the distribution function remains nearly Maxwellian, permitting a $\delta f$ rather than full-$f$ approach
and linearized treatment of collisions, and the $\vect{E}\times\vect{B}$-drift nonlinearity also becomes negligible.
Any more ambitious effort to analyze a pedestal with $r_T \sim \rhop$ for finite aspect ratio will likely need to retain both these nonlinearities,
necessitating complicated codes, which our results may be used to benchmark.
We also present a new numerical continuum approach to computing these global neoclassical effects
in the weak-$\Ti'$ limit, including the exact linearized \FP collision operator. We
exploit the success of local continuum neoclassical codes by making such a local code the inner step of an iteration
loop for the global calculation.

Several local neoclassical codes have been developed \cite{Sauter0,NCLASS,DarinNotch,NEO1,WongChan,Lyons},
and other numerical efforts have
computed nonlocal
neoclassical effects in transport barriers using the particle-in-cell (PIC)
approach
\cite{Lin1,Wang1,CSChang1,ORB5}.
Since PIC and continuum codes have differing treatments of collisions and boundary conditions and differing numerical resolution challenges, it is good practice to develop both approaches to ensure they yield the same physical results.
Some neoclassical investigations have been made in global continuum codes \cite{Xu1,COGENT},
but these codes are ultimately designed for turbulence studies, and very different algorithms have
been used than the one we describe.
Some analytic results are available\cite{GrishaNeo,GrishaPRL,Istvan}, but only in restricted limits of aspect ratio and collisionality,
where simplified collision models are expected to be valid.

Throughout our analysis we assume $\Bp \ll B$, where $B=| \vect{B}|$ is the magnetic field strength and $\Bp$ is the poloidal field,
implying a scale separation between $\rhop$ and the gyroradius $\rho$.
Without this approximation,
a gyrokinetic
rather than drift-kinetic treatment would be necessary, including changes to the collision operator\cite{CattoTsang, BoDarin}


In conventional neoclassical theory, the ion distribution function is expanded as $\fii = \fMi + f_1$
where $\fMi \gg f_1$, and
$\fMi$ is a Maxwellian with constant density $\ni$ and temperature $\Ti$ on each flux surface.
The drift-kinetic equation is then solved for $f_1$, with the result that $f_1$ includes a term
$-(I\vpar/\Omega)\partial \fMi/\partial \psi$.
Here, $I$ equals the major radius $R$ times the toroidal field $B_{\mathrm{tor}}$,
$\Omega=ZeB/\mi c$, $Z$ is the ion charge in units of the proton charge $e$,
$\mi$ is the ion mass,
 $c$ is the speed of light, and $2\pi\psi$ is the poloidal flux.
The derivative is carried out at fixed total unperturbed energy $W_0 = \mi\vv^2/2 + Ze\Phio$,
where $\Phio=\left< \Phi \right>$ is the flux-surface average of the electrostatic potential $\Phi$.
We may estimate $\partial \fMi/\partial\psi \sim \fMi/(R\Bp r_\bot)$,
so $f_1 \sim (\rhop/r_\bot)\fMi$
where
$\rhop = B\vi/(\Bp \Omega)$ is the poloidal ion gyroradius, and $\vi = \sqrt{2\Ti/\mi}$ is the ion thermal speed.
In a pedestal, since
$\rhop/r_\bot \sim 1$, then $f_1 \sim \fMi$, so conventional neoclassical results are no longer valid.

However, a more precise analysis reveals\cite{GrishaNeo} a regime
in which the near-Maxwellian assumption is still appropriate.
Writing
$\fMi = \eta(\psi) \{\mi/[2\pi \Ti(\psi)]\}^{3/2} \exp( -W_0 / \Ti(\psi))$, where $\eta(\psi) = \ni(\psi) \exp\left( Z e \Phio(\psi)/\Ti(\psi)\right)$,
the derivative $(\partial \fMi/\partial\psi)_{W_0}$ that determines the magnitude of $f_1$ is
\begin{equation}
\frac{\partial \fMi}{\partial\psi}
= \left[\frac{1}{\eta} \frac{d \eta}{d\psi} + \left(W_0-\frac{3}{2}\right) \frac{1}{\Ti} \frac{d\Ti}{d\psi}  \right] \fMi.
\label{eq:dfmdpsiEta}
\end{equation}
The magnitude of $\partial \fMi/\partial\psi$ is evidently determined by
$r_T$ and $r_\eta$, the scale-lengths of $\Ti$ and $\eta$, but not directly by $r_n$, the scale-length of density.
Observing $r_{\eta}^{-1} = r_n^{-1} - Ze\Phio'/\Ti + Ze\Phio/(\Ti r_T)$,
$f_1 /\fMi$ may be small even when $r_n \sim \rhop$ as long as $r_T$ and $r_\eta$ are $\gg \rhop$.
Such is the case when $d\Phio/d\psi \simeq \Ti (Ze\ni)^{-1} d\ni/d\psi$ so the ions are electrostatically
confined.

We consider this ``weak-$\Ti'$ pedestal" regime for the rest of the analysis:
$r_n \sim \rhop$ but $\delta \ll 1$ where $\delta = \rhop / r_T$ is
the basic expansion parameter, and $ \rhop / r_\eta \sim \delta$.  (The electron temperature $\Te$
is free to vary on the $\rhop$ scale.)
This ordering, also considered in Ref. \onlinecite{GrishaNeo},
is useful in part because
the collision operator may be linearized.
Also, as we will show, the poloidal electric field decouples from the kinetic equation,
so the equation becomes linear in $f_1$.
For $r_T \sim \rhop$ and/or $r_\eta \sim \rhop$,
the full bilinear collision operator must be used and a full-$f$ nonlinear kinetic equation must be solved,
including the electric field nonlinearity.
Notice $r_\eta \ll \rhop$ implies $(Ze/\Ti)d\Phio/d\psi \sim 1/(R\Bp\rhop)$
and so $Z e\Phio/\Ti\sim 1$.
As a result, the term $\vE\cdot\nabla \fii$ in the kinetic equation,
neglected in conventional theory, becomes comparable in magnitude
to the $\vpar\gradpar \fii$ term.
Thus, even though the weak-$\Ti'$ ordering permits $f_1 \ll \fMi$, conventional neoclassical
results still must be modified.
As $\Bp \ll B$, the $\vect{E}\times\vect{B}$ drift $\vE$ satisfies $|\vE| \ll \vi$ so centrifugal effects may be neglected.

We begin with the ion drift-kinetic equation\cite{Hazeltine}
\begin{equation}
(\vpar\vect{b} + \vd)\cdot(\nabla \fii)_{\mu,W} = \Cl\{\fii\}+S
\label{eq:fullFDKE}
\end{equation}
where the gradients hold fixed $\mu=\mi \vv^2/(2B)$ and $W=\mi \vv^2/2+Ze\Phi$ (now including
$\Phi$, not just $\Phio$), $\Cl$ is the ion-ion collision operator linearized about $\fMi$,
and $S$ represents any sources/sinks.
We take $\vd = (\vpar/\Omega)\left.\nabla\right|_W\times(\vpar \vect{b})$
(which includes $\vE$.)

Now change from $W$ to $W_0=W-Ze\Phi_1$ as an independent variable, where $\Phi_1 = \Phi-\Phio$.
We assume $\Phi_1 \sim \delta \Phio$ and $\partial \Phi_1/\partial \psi \sim \delta d\Phio/d\psi$,
and we will show in a moment these orderings are self-consistent.
Then defining $g$ by
\begin{equation}
\fii = \fMi -(Ze\Phi_1/\Ti)\fMi - (I\vpar/\Omega)\partial \fMi/\partial\psi +g,
\label{eq:gDef}
\end{equation}
(\ref{eq:fullFDKE}) may be written
\begin{equation}
\left( \vpar \vect{b} + \vd \right)\cdot(\nabla g)_{\mu,W_0} - \Cl\{g\} =
 \CI+S
\label{eq:g}
\end{equation}
where $\CI = \Cl\{(I\vpar/\Omega)\partial \fMi/\partial\psi\}$ is the inhomogeneity,
the independent variable is now $W_0$, and terms small in $\delta$
have been dropped.  The contribution from $\Phi_1$ to $\vd\cdot\nabla\theta$ is $O(\delta)$
smaller than the $\Phio$ contribution, and
$(\vE\cdot\nabla\psi)/(\vm\cdot\nabla\psi) \sim Ze\Phi_1/\Ti \sim \delta$ where $\vm=\vd-\vE$ is the magnetic drift,
so we may approximate $\vd$ in (\ref{eq:g}) with the leading-order drift $\vdo = \vm+\vEo$
where $\vEo=(c/B^2)\vect{B}\times\nabla\Phio$.
Then (\ref{eq:g}) is completely linear.
To evaluate $\Phi_1$ we may use the electron density $ \nee + (e\Phi_1/\Te)\nee$ with quasineutrality
to find
$e\Phi_1/\Ti  = \left(\Ti/\Te+Z \right)^{-1}\ni^{-1}\int d^3\vv\, g.$
Hence, as $g \sim \delta \fMi$, our assumed ordering for $\Phi_1$ is self-consistent.
Using
$\Cl\{\vpar \fMi\}=0$,
the only gradient surviving in $\CI$
is $dT/d\psi$.
While the independence of $g$ from $d\ni/d\psi$ and $d\Phio/d\psi$
was  known previously for the local case,
the persistence of this property in the weak-$\Ti'$ pedestal case is noteworthy\cite{GrishaNeo}.

One crucial difference between the local and global analyses is that the flow may vary
over a flux surface in different ways.
First consider the parallel ion flow $\ni \Vipar=\int d^3\vv \, \vpar \fii$:
\begin{equation}
\ni\Vipar
= -\frac{cI}{ZeB} \left( \frac{d \ppi}{d\psi} + Ze\ni \frac{d\Phio}{d\psi} - \kV \frac{B^2}{\left< B^2\right>} \ni\frac{d\Ti}{d\psi}\right)
\label{eq:kVdef}
\end{equation}
where $\kV = Ze\left<B^2\right> (cI\ni B\,d\Ti/d\psi)^{-1}\int d^3\vv\,\vpar g$ is dimensionless.
We have exploited the aforementioned fact $g \propto d\Ti/d\psi$.
In the conventional ordering, $\kV$ is also the coefficient of the poloidal flow $V_{\theta}$:
forming the appropriate linear
combination of (\ref{eq:kVdef}) with the perpendicular diamagnetic and $\vect{E}\times\vect{B}$ flows,
$V_{\theta} = \vect{V}\cdot \vect{e}_\theta = \kV c I B_\theta \left( Z e \left< B^2\right>\right)^{-1} d\Ti/d\psi$
where $\vect{e}_\theta=(\nabla \zeta\times\nabla\psi)/|\nabla\zeta\times\nabla\psi|$
and $\Bp=\vect{B}\cdot\vect{e}_\theta$.

Applying $\int d^3\vv = 2\pi \mi^{-2} \sum_\sigma \sigma \int dW \int d\mu (B/\vpar)$ to
(\ref{eq:g}), where $\sigma=\sgn(\vpar)$,
the resulting mass conservation equation (ignoring $S$) is
\begin{equation}
\frac{\partial}{\partial\theta}   \int d^3\vv(\vpar+u) \frac{g}{B}  - \frac{\partial}{\partial\psi}   \int d^3\vv \frac{g\vm\cdot\nabla\psi}{\vect{B}\cdot\nabla\theta} = 0
\label{eq:globalConservation}
\end{equation}
where $u = (\vdo\cdot\nabla\theta)/\gradpar\theta \approx (cI/B)d\Phio/d\psi$
is comparable in magnitude to $\vpar$.
In the local case,
where $\vd\cdot\nabla g$ is neglected in  (\ref{eq:g}),
only the first term in (\ref{eq:globalConservation})
 ($\propto\vpar$) arises, implying
$\int d^3\vv\,\vpar g \propto B$ and $\partial \kV/\partial\theta=0$.
This is the origin of the well known conventional
result that
$\kV$ is constant on a flux surface.  However, in the global
case,
the strong
poloidal drift and $\rhop$-scale radial variation
drive poloidal variation in $\kV$.

The total flow remains divergence-free in a fluid picture:
$\nabla\cdot \vect{\Gamma}=0$  where
$\vect{\Gamma} =  \Gamma_{||} \vect{b} + \vect{\Gamma}_E + \vect{\Gamma}_{\mathrm{dia}}$,
$\Gamma_{||} = \int d^3\vv\,\vpar \fii$,
$\vect{\Gamma_E} = \ni \vEone + \int d^3\vv\, \fii\vEo$ contains the first
two orders of the $\vect{E}\times\vect{B}$ flux,  $\vEone=(c/B^2)\vect{B}\times\nabla\Phi_1$,
$\vect{\Gamma}_{\mathrm{dia}}=c(ZeB^2)^{-1}\vect{B}\times\nabla\cdot\tens{\Pi}$
contains the first two orders of the diamagnetic flow,
$\tens{\Pi}=p_{\bot}(\tens{I}-\vect{b}\vect{b})+p_{||}\vect{b}\vect{b}$,
$p_{\bot}=m\int d^3\vv\,\fii \vv_\bot^2/2$, and $p_{||}=m\int d^3\vv\, \fii \vpar^2$.
To prove $\nabla\cdot\vect{\Gamma}=0$ from (\ref{eq:globalConservation}),
(\ref{eq:gDef}) and $\nabla \ni \approx -(Ze\ni/\Ti)\nabla \Phio$ are applied, along with
$\int d^3\vv\, \fii\vm = \vect{\Gamma}_{\mathrm{dia} } + \nabla\times\vect{M} + \vect{\Gamma}_{\mathrm{f}}$
(true for any $\fii$). Here
$\vect{M}=\vect{b}cp_{\bot}/(ZeB)$, and we will neglect the
$O(\beta \delta)$ parallel flow correction $\vect{\Gamma}_{\mathrm{f}}=(p_{||}-p_{\bot })\vect{b}\vect{b}\cdot\nabla\times\vect{b}$ (which disappears when a more accurate $\vm$ is used.)
As before, $\ni(\psi)=\int d^3\vv\, \fMi$ includes only the leading-order density. We have needed to
keep terms of two orders in both $\vect{\Gamma}_E$ and $\vect{\Gamma}_{\mathrm{dia}}$ because the $\vect{E}\times\vect{B}$ and diamagnetic flows cancel to leading order in our ordering.
And, though $\vect{\Gamma}_E \approx \ni \vEo$ and $\tens{\Pi} \approx \ppi \tens{I}$, the radial derivative in $\nabla\cdot\vect{\Gamma}$ means the
next-order corrections to these terms must be retained to accurately compute $\nabla\cdot\vect{\Gamma}$.
The poloidal fluid velocity is defined by
$V_\theta=\vect{\Gamma}\cdot\vect{e}_\theta/\ni$.
It can be shown
that to leading order in $\delta$,
\begin{equation}
V_\theta \approx \frac{\Bp}{\ni B}\left[ \int d^3\vv\left(\vpar+\frac{cI}{B}\frac{d\Phio}{d\psi}\right)g  + \frac{I}{\Omega} \frac{\partial}{\partial \psi} \int d^3\vv\frac{\vv_{\bot}^2}{2}g\right].
\end{equation}
In the local case, the $\vpar$ term dominates, so $V_\theta \propto \kV$.
In the global case, $V_\theta$ remains proportional to $d\Ti/d\psi$, but $V_\theta$ is no longer $\propto \kV$.
A normalized poloidal flow may be defined by
\begin{equation}
k_\theta=V_\theta Z e \left< B^2\right> / (c I B_\theta\,d\Ti/d\psi)
\end{equation}
so $k_\theta \to \kV$ in the local limit.
That $\kV \ne k_\theta$ in the pedestal is a central new result of this work.


As a result of these flow modifications, the current also changes.
We write the electron distribution $\fe = \fMe \exp(e\Phi_1/\Te)+h$ (in the gauge
$E_{||}=B\left<E_{||}B\right>/\left<B^2\right> -\gradpar\Phi$) with $\fMe$ the electron Maxwellian.
Keeping $O(1)$ and $O(\delta)$ terms in the electron
kinetic equation with independent variable $w_0=\me\vv^2/2-e\Phio$, (assuming $\sqrt{\me/\mi} \ll \delta$,)
\begin{eqnarray}
\vpar\gradpar h + (\vme+\vEone)\cdot\nabla\fMe
+e\Phi_1 \vme\cdot\nabla\frac{\fMe}{\Te}
+e\vpar\frac{\partial h}{\partial w_0} \gradpar\Phi_1
 + \frac{e \vpar \left<\Epar B\right> B}{\Te\left<B^2\right>}\fMe = \Ce. \nonumber \\
\label{eq:electronKineticEquation}
\end{eqnarray}
Here $\vme$ is the electron magnetic drift, $\Ce = \Cee + \Cei$ is the electron collision operator,
$\Cei \approx \nuei L\{h\} + \fMe\nuei\me\vpar\Vipar/\Te$, $L = (1/2) (\partial/\partial\xi)(1-\xi^2)(\partial/\partial\xi)$,
and $\xi=\vpar/\vv$.
Expanding $h=h_0+h_1$ with $h_1/h_0\sim\delta$, the leading order solution of (\ref{eq:electronKineticEquation}) (i.e.
neglecting $\Phi_1$ and $d\Ti/d\psi$ terms) gives $h_0$
representing the usual \PS and bootstrap currents but without the $d\Ti/d\psi$ contribution.
At next order, $h_1=\fMe\me \vpar\Vipar/\Te-cI\ni h_{\Ti}(d\Ti/d\psi)/e-\rho_0 cI^2 (d\nee/d\psi)(d\Ti/d\psi)h_\Phi/e$  where
$\rho_0=\vi \mi c/(Ze\Bav)$, $\Bav^2=\left<B^2\right>$, and
$h_{\Ti}$ and $h_\Phi$ are the solutions of
\begin{eqnarray}
D h_{\Ti} &=& \fMe \me (\nee \Te)^{-1} \left< B^2\right>^{-1} \vpar \gradpar \left( \vpar B \kV \right),
\label{eq:h}
\\
 a D h_\Phi &=&
\vEone\cdot\nabla \fMe
+e\Phi_1\vme\cdot\nabla\frac{\fMe}{\Te}
+e\vpar \frac{\partial h_0}{\partial w_0} \gradpar\Phi_1
\nonumber 
\end{eqnarray}
with $D=\vpar \gradpar - \Cee - \nuei L$ and $a=\rho_0 c I^2 e^{-1} (d\nee/d\psi)(d\Ti/d\psi)$.
Applying $\int d^3\vv$ to (\ref{eq:h}),
$\int d^3\vv\,\vpar h_{\Ti} = \alpha_{\Ti} B + \kV B/\left< B^2\right>$ and $\int d^3\vv\,\vpar h_\Phi=\alpha_\Phi B - n_g/(Z B)$
where $\alpha_{\Ti}$ and $\alpha_{\Phi}$ are flux functions,
$n_g=\Ti (\rho_0 I \ni\, d\Ti/d\psi)^{-1}\int d^3\vv\, g$ is the $O(1)$ normalized density perturbation, and we have invoked quasineutrality.
Then forming $j_{||}=e\int d^3\vv(Z\fii-\fee)\vpar$,
\begin{eqnarray}
\jpar = \frac{cI}{B} \frac{dp}{d\psi} \left( \frac{B^2}{\left< B^2\right>}-1\right) + \frac{cI \nee B}{Z \left< B^2\right>} \frac{d\Ti}{d\psi}\left( \kV - \frac{\left< B^2 \kV\right>}{\left< B^2 \right>}\right) \nonumber \\
+\frac{\rho_0 c I^2}{Z} \frac{d\nee}{d\psi} \frac{d\Ti}{d\psi} \left( \frac{\left< n_g\right> B}{\left< B^2\right>} -  \frac{n_g}{B} \right)
+ \frac{\left< \jpar B\right> B}{\left< B^2 \right>}
\label{eq:currentVariation}
\end{eqnarray}
where $p=\pe+\ppi$.
The $dp/d\psi$ and $\left< \jpar B \right>$ terms arise in the local case; the former is the standard
\PS current, and the latter is the Ohmic and bootstrap contribution.
The $\kV$ and $n_g$ terms however have not been reported previously.
Curiously, the $n_g$ term is quadratic in the gradients.
The Ohmic and bootstrap contribution is
\begin{equation}
\left< \jpar B\right>
= \sigmaneo \left< E_{||}B\right> -cI\pe \left(  \frac{\Lone}{\pe}\frac{dp}{d\psi}
+ \frac{\Ltwo}{\Te}\frac{d\Te}{d\psi}  - \frac{\LTi}{Z \Te} \frac{d\Ti}{d\psi}
-\frac{\LnT\rho_0 I}{\nee\Te}\frac{d\nee}{d\psi}\frac{d\Ti}{d\psi}
\right)
\label{eq:jbs}
\end{equation}
using notation of Ref. \onlinecite{Sauter},
where $\sigmaneo$, $\Lone$, and $\Ltwo$ are calculated in the standard way, and
$\LTi = \left< B \int d^3\vv\; \vpar h_{\Ti}\right>$ and
$\LnT = \left< B \int d^3\vv\; \vpar h_{\Phi}\right>$ are new dimensionless coefficients.
In the local case of constant $\kV$, (\ref{eq:h}) shows $\LTi \propto \kV$.
However, to determine $\LTi$ in the global case, (\ref{eq:h}) must be solved accounting for the poloidal
variation of $\kV$.
As with the flow, the total current is divergence-free: (\ref{eq:currentVariation}), (\ref{eq:globalConservation}), and quasineutrality
imply (after some algebra) $0=\nabla\cdot\vect{j}=\nabla\cdot(\jpar \vect{b}+cB^{-2}\vect{B}\times\nabla\cdot\tens{\Pi}_{\Sigma})$,
where the ion plus electron stress $\tens{\Pi}_{\Sigma}$ is computed from (\ref{eq:gDef}) and $\fee\approx \fMe(1+e\Phi_1/\Te)$.
The new $\kV$ and $n_g$ terms in (\ref{eq:currentVariation}) arise for the same reason as the usual \PS current: a parallel return current must flow
to maintain $\nabla\cdot\vect{j}=0$ given the perpendicular diamagnetic current. In the pedestal, the pressure variation on a flux surface
becomes sufficient to modify this diamagnetic current.

\begin{figure}
\includegraphics{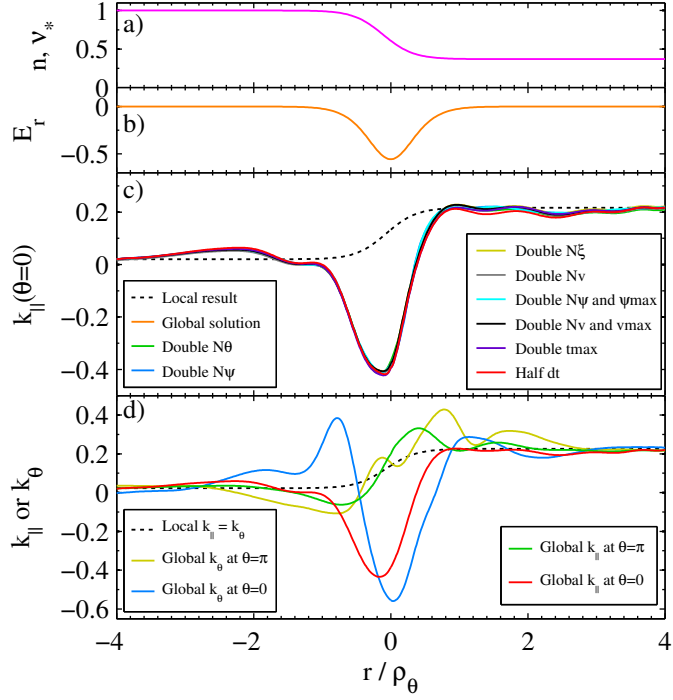}
\caption{(Color online)
a) Equilibrium density, normalized to its value at the left boundary.  As $\nu_*$ happens to be 1 at this boundary
and $\Ti \approx$ constant over the domain, this plot also gives the $\nu_*$ profile.
b) Normalized radial electric field $-c I (\vi B_0)^{-1} d\Phio/d\psi$.
c) The $d\Ti/d\psi$-driven parallel flow $\kV$ computed
in the local approximation (dashed curve) differs from the global result (nearly indistinguishable solid curves) in the pedestal.
The global code is well converged, demonstrated by changing each resolution parameter by $2\times$.
d) Normalized poloidal flow $k_\theta$ and $\kV$, evaluated at the outboard ($\theta=0$) and inboard ($\theta=\pi$) midplanes.
\label{fig:globalKVProfile}}
\end{figure}

We now discuss our numerical method for solving the pedestal ion kinetic equation.
The radial domain is an annulus containing the pedestal, several $\rhop$ wide.
As $r_\eta, r_T \gg \rhop$, we take $\eta$ and $\Ti$ constant over this domain for simplicity.
Also, radial variation of $I$, $B$, and $\gradpar \theta$ is neglected.
We specify $\ni(\psi)$, which determines $\Phio=(Ze)^{-1}\Ti \ln(\eta/\ni)$.
On either end of the radial domain,  $\ni(\psi)$ and $\Phio(\psi)$ are uniform for several $\rhop$, as in
figure \ref{fig:globalKVProfile}.a-b, allowing local solutions to be used for inhomogeneous Dirichlet radial boundary conditions.
We discretize in the variables $(\psi,\theta,  \vv,\xi)$.

To solve (\ref{eq:g}),  $\partial g/\partial t$ is first added to the left-hand side,
and with the local solution as an initial condition, $g$ is evolved to equilibrium using
the following operator-splitting method.
Consider the successive backwards-Euler time steps
\begin{eqnarray}
\left[g_{t+(1/2)} - g_t\right] / \Delta t + \Knl\{g_{t+(1/2)}\} &=& 0,
\label{eq:step1}
\\
\left[ g_{t+1} - g_{t+(1/2)} \right] / \Delta t + \Kloc\{g_{t+1}\} &=& \CI + S,
\label{eq:step2}
\end{eqnarray}
where $\Knl = (\vm\cdot\nabla\psi) (\partial/\partial\psi)_{\vv,\xi}$
is the ``nonlocal" term,
and in $\Kloc = (\vpar\vect{b}+\vdo)\cdot(\nabla)_{\mu,W_0} - \Cl - \Knl$, $\psi$ is only a parameter.
In the sum (\ref{eq:step1})$+$(\ref{eq:step2}), $g_{t+(1/2)}/\Delta t$ cancels, leaving an equation equivalent to first order in $\Delta t$ to a
step with the complete operator $\Knl + \Kloc$.
However, (\ref{eq:step1}) and (\ref{eq:step2}) are much easier than a step with the total operator because the dimensionality is reduced.
Also notice the local and nonlocal operators at each grid point need only be $LU$-factorized once, with the $L$ and $U$ factors reused at each
time step for rapid implicit solves.

Our approach to implementing the full \FP field operator, similar to the local code in Ref. \onlinecite{Lyons}, is to treat the Rosenbluth potentials\cite{RMJ}
$H$ and $G$ as unknown fields along with $g$, and to solve a block linear system for three simultaneous equations: (\ref{eq:step2}), $\gradv^2 H = -4\pi g$, and $\gradv^2 G = 2H$,
with $\gradv^2$ the velocity-space Laplacian.
Our local solver has been successfully benchmarked against many analytic formula
and against results of another Fokker-Planck code\cite{WongChan}.
More details of the numerical implementation will be described in a forthcoming publication.

The heat fluxes at the two radial boundaries are different due to the different densities, so heat will accumulate in the simulation domain,
precluding equilibrium unless an appropriate heat sink is present.
In a real pedestal, there will be a divergence of the \emph{turbulent} fluxes,
which could act as this sink in the long-wavelength (drift-kinetic) equation
we simulate here.  Determining the phase-space structure of this sink
from first principles is beyond the scope of this work, so we use
$S = -\gamma \left<g(\xi) + g(-\xi)\right>$
for constant $\gamma$, resembling the sink in  Ref. \onlinecite{GENESource} for global
$\delta f$ gyrokinetic codes.  Varying $\gamma$ by several orders of magnitude
or using different forms of $S$ cause little change to the results.

\begin{figure}
\includegraphics{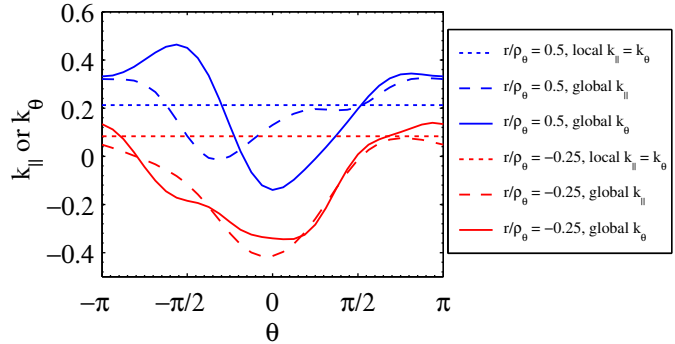}
\caption{(Color online)
Poloidal variation of the parallel flow coefficient $\kV$ and normalized poloidal flow $k_\theta$
at two radial locations straddling the pedestal.
\label{fig:spatialKVVariation}}
\end{figure}

Figures \ref{fig:globalKVProfile}-\ref{fig:spatialKVVariation} show results of the global calculation
for a pedestal with $\epsilon=0.3$, $B=B_0/[1+\epsilon \cos(\theta)]$, and $\gradpar \theta=$constant.
The density decreases by $3\times$ from the top of the pedestal to the bottom,
varying $\nustar = \nu_{\mathrm{ii}}/(\epsilon^{3/2}\vi\gradpar\theta)$ from $1 - 0.3$.  The electric field profile consistent with this density profile for $r_\eta \gg \rhop$ is shown in figure
\ref{fig:globalKVProfile}.b.  The electric field reaches a maximum magnitude of $\approx -0.5 \vi \Bp/c$ in the middle of the pedestal.
In these plots, the radial coordinate $r/\rhop$ is defined by $r/\rhop = ZeB_0 (\mi c\vi I)^{-1} \psi$ where $B_0$ is the toroidal field on axis;
 $r=0$ is an arbitrary minor radius, not the magnetic axis.
For the sink, $\gamma=0.1 \omegat$ where  $\omegat=\vi \gradpar\theta$ is the ion transit frequency.
The simulation is run to $t=100/\omegat$, since doubling this duration produces negligible difference in the results.
Figures \ref{fig:globalKVProfile}.c-d and \ref{fig:spatialKVVariation} show the parallel flow coefficient $\kV$ and the normalized poloidal flow $k_\theta$.
For comparison, the local $\kV=k_\theta$ is also shown,
computed at each $r$ by numerical solution of (\ref{eq:g}) without $\vd$ or $S$.
Even in the local case, $\kV$ and $k_\theta$ vary slightly across the pedestal due to the change in collisionality.
Outside of the pedestal, as expected, $\kV$ and $k_\theta$ computed by the global code
are equal, constant on each flux surface, and unchanged from the local (conventional) result.
Inside the pedestal, $\kV$ and $k_\theta$  differ from the local result, and both coefficients vary poloidally and change sign.
The most dramatic change is a well in $\kV$  and $k_\theta$ at the outboard midplane.
Although the distribution for an up-down symmetric $B$ field has the symmetry $g(-\theta,-\vpar) = -g(\theta,\vpar)$ in the local case,
in the global case the drift
 terms in the kinetic equation break this symmetry, so the global curves in Figure \ref{fig:spatialKVVariation}
lack definite $\theta$ parity.
To verify mass conservation, the $\vpar$, $u$, and $\vm$ terms in (\ref{eq:globalConservation})
were each independently computed from $g$, and it was verified that the result indeed summed to zero.

To conclude,
in this work we have
demonstrated
an extension of neoclassical calculations to a density pedestal with $r_n \sim \rhop$ but $r_T \gg \rhop$, retaining effects of finite orbit width,
collisionality, and aspect ratio.
The kinetic equation remains linear, and a $\delta f$ approach is possible.
A numerical scheme was illustrated, demonstrating convergence on a laptop for experimentally relevant parameters.
The Rosenbluth potentials are solved for along with the distribution function at each step, allowing use of the full linearized \FP collision operator.

The analytic and numerical calculations show that in a pedestal, the plasma flow can differ significantly from the conventional prediction.
While the poloidal flow is $\propto B_\theta$ in the core, the same is not generally true
in the pedestal, and while the numerical coefficients in the parallel and poloidal flow are identical in conventional theory, in the pedestal these coefficients $\kV$ and $k_\theta$
are generally different.
These modifications may be important for comparisons of experimental pedestal flows to theory.\cite{Kenny1}
Two new contributions to mass conservation become important which are normally neglected: $\vect{E}\times\vect{B}$ motion of the perturbed density, and diamagnetic flow of the pressure perturbation.
In general, the poloidal flow and $d\Ti/d\psi$ component of the parallel flow can differ in both magnitude and sign relative to local theory, as shown in the figures.

Associated with the modification to the flow, the usual division of the parallel current into \PS and Ohmic-bootstrap components is changed (Eq. (\ref{eq:currentVariation})), and
the $d\Ti/d\psi$ contribution to the bootstrap current is altered.  In the weak-$\Ti'$ orderings used here, the associated terms are necessarily smaller than adjacent $dp/d\psi$ terms.  However, analogous modifications to the current would presumably occur in a full-$f$ calculation when $r_T \sim \rhop$, giving order-unity departures from local theory in that case.


%
%

%


We are grateful to Peter Catto and Felix Parra for enlightening discussions and for reading the manuscript.
We also thank S. Kai Wong and Vincent Chan for assistance with benchmarking our local code to that of Ref. \onlinecite{WongChan}.
This work was supported by the
Fusion Energy Postdoctoral Research Program
administered by the Oak Ridge Institute for Science and Education.



%

\end{document}